\documentclass{vgtc}                          % final (conference style)
\ifpdf%                                % if we use pdflatex
  \pdfoutput=1\relax                   % create PDFs from pdfLaTeX
  \usepackage{graphicx}                % allow us to embed graphics files
  \DeclareGraphicsExtensions{.pdf,.png,.jpg,.jpeg} % for pdflatex we expect .pdf, .png, or .jpg files
\else%                                 % else we use pure latex
  \usepackage{graphicx}                % allow us to embed graphics files
  \DeclareGraphicsExtensions{.eps}     % for pure latex we expect eps files
\fi%

\graphicspath{{figures/}{pictures/}{images/}{./}} % where to search for the images

\usepackage{microtype}                 % use micro-typography (slightly more compact, better to read)
\PassOptionsToPackage{warn}{textcomp}  % to address font issues with \textrightarrow
\usepackage{textcomp}                  % use better special symbols
\usepackage{mathptmx}                  % use matching math font
\usepackage{times}                     % we use Times as the main font
\usepackage{cite}                      % needed to automatically sort the references
\usepackage{tabu}                      % only used for the table example
\usepackage{booktabs}                  % only used for the table example

\usepackage[official]{eurosym}
\usepackage{multirow}
\usepackage{flushend}

\usepackage{xcolor}

\onlineid{0}

\vgtccategory{Application papers}
\vgtcinsertpkg

\title{Effects of Self-Avatar and Gaze on Avoidance Movement Behavior\\
~\\ {\large To Appear in IEEE VR 2019, the 26th IEEE Conference on Virtual Reality and 3D User Interfaces}
}

\author{Christos Mousas\thanks{e-mail: cmousas@purdue.edu}\\ %
\parbox{1in}{\scriptsize \centering Purdue University}
\and Alexandros Koilias\thanks{e-mail: ctd17008@aegean.gr}\\ %
\parbox{1.2in}{\scriptsize \centering University of the Aegean}
\and Dimitris Anastasiou\thanks{e-mail: anastasiou@siu.edu}\\ %
\parbox{1.9in}{\scriptsize \centering Southern Illinois University Carbondale }
\\
\\
\and Banafsheh Rekabdar\thanks{e-mail: brekabdar@cs.siu.edu}\\ %
\parbox{1.9in}{\scriptsize \centering Southern Illinois University Carbondale}
\and Christos-Nikolaos Anagnostopoulos\thanks{e-mail: canag@ct.aegean.gr}\\ %
\parbox{1.2in}{\scriptsize \centering University of the Aegean}
}

\teaser{
\includegraphics[width=\textwidth]{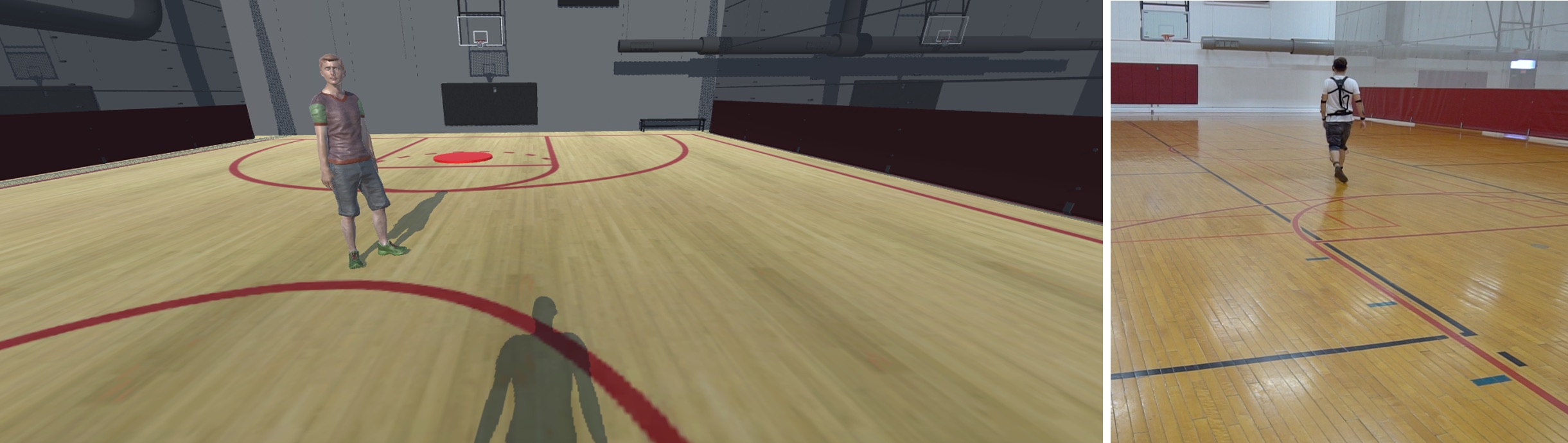}
\caption{A participant walking during the experiment and the content that the participants are seeing through the HMD during the Self-Avatar LookAt condition of the experiment.}
\label{fig1}
}

\abstract{
The present study investigates users' movement behavior in a virtual environment when they attempted to avoid a virtual character. At each iteration of the experiment, four conditions (Self-Avatar LookAt, No Self-Avatar LookAt, Self-Avatar No LookAt, and No Self-Avatar No LookAt) were applied to examine users' movement behavior based on kinematic measures. During the experiment, 52 participants were asked to walk from a starting position to a target position. A virtual character was placed at the midpoint. Participants were asked to wear a head-mounted display throughout the task, and their locomotion was captured using a motion capture suit. We analyzed the captured trajectories of the participants' routes on four kinematic measures to explore whether the four experimental conditions influenced the paths they took. The results indicated that the Self-Avatar LookAt condition affected the path the participants chose more significantly than the other three conditions in terms of length, duration, and deviation, but not in terms of speed. Overall, the length and duration of the task, as well as the deviation of the trajectory from the straight line, were greater when a self-avatar represented participants. An additional effect on kinematic measures was found in the LookAt (Gaze) conditions. Implications for future research are discussed.
} % end of abstract

\CCScatlist{
\CCScatTwelve{Human-centered computing}{Human computer interaction (HCI)}{Interaction paradigms}{Virtual reality};
}

\begin{document}


\firstsection{Introduction}
\label{sec1}
\maketitle

%% \section{Introduction} %for journal use above \firstsection{..} instead
Motion-sensing and displaying technologies allow users to dive into virtual worlds and interact with virtual content. Low-cost motion capture systems and head-mounted displays (HMDs) have become readily accessible to the public in recent years. The availability of these technologies makes it possible to acquire the necessary devices and applications to experience an immersive virtual world from the comfort of one's living room. In immersive environments, an important issue that needs further investigation is the interaction with virtual characters when users perform locomotive tasks. Understanding such interactions can have practical applications in home entertainment and clinical practice (e.g., preventive or remedial treatment for students with autism spectrum disorder).

In the current experiment, we asked participants to perform a walking motion (see Figure \ref{fig1}) from a starting position to a target position, and a virtual character was placed at the midway point. Each time the participants performed the walking task, four conditions of the experiment were applied by combining participants' representation (Self-Avatar versus No Self-Avatar) and the gaze of the virtual character that they were instructed to avoid (LookAt versus No LookAt). It should be noted that the position of the character was fixed (not moving in the opposite direction), and the character was assigned to have an idle motion. During the experiment, the participants' full-body motion was captured. The captured motion sequences were later analyzed to examine whether the applied conditions had an impact on the paths of participants when avoiding the virtual character.

In the real world scenarios, the literature is not conclusive of how humans adjust their movement behavior in the presence of a non-locomotive human \cite{ref66}. Moussaid et al. \cite{ref80} reported that when avoiding a non-locomotive human, people simply change their movement direction to avoid a possible collision. Huber et al. \cite{ref66} found that speed adjustments of walkers were made only in scenarios where the interferer was crossing at acute angles. Additionally, other studies found that the avoidance behavior of humans is highly dependent on the environmental constraints and the dynamics (e.g., locomotive versus non-locomotive) of the avoidance human \cite{ref68, ref81}.
 
In virtual reality environments, a number of studies have been developed to understand the collision avoidance, gaze interaction, and paths that participants follow when facing virtual characters. The current experiment advances methods for analyzing participants' locomotion using kinematic measures \cite{ref4, ref42} and uses the avoidance behavior of participants as a method to study how the Self-Avatar and LookAt dimensions influence the way that participants interact with virtual characters.

The main goal of the present study, which conducted in a virtual reality environment, is to investigate the effects of self-avatar and the gaze of virtual character on human movement behavior during a collision avoidance task by using kinematic measures. Specifically, the research questions of this study are as follows:
\begin{itemize}
\item \textbf{RQ 1:} Does the length of the participants' trajectory differ in each of the four experimental conditions?
\item \textbf{RQ 2:} Does the duration of the task (walk from the start to the target position) differ across the four experimental conditions?
\item \textbf{RQ 3:} Does the average walking speed of the participants differ across the four experimental conditions?
\item \textbf{RQ 4:} Does the participants' deviation from the straight line in the captured trajectory differ across the four experimental conditions?
\end{itemize}

\section{Related Work}
\label{sec2}
This section presents related work on collision avoidance, gaze interaction, and motion analysis.

\subsection{Collision Avoidance Behavior}
\label{sec21}
During everyday life, when walking, people try to maintain a secure distance from other humans. This is achieved by adapting motion. Coren \cite{ref13}, working on laterality, found that the interaction between a walker and the environment can be modeled as a dynamic system. From a global perspective \cite{ref14, ref64}, heading direction changes based on the distance, the angle between the walker, the target positions, and the obstacles that are located in the environment. From a local perspective \cite{ref16, ref15}, a walker will avoid an obstacle or a virtual character using anticipatory locomotor adjustment behavior, which means that the width of steps is adapted before the avoidance behavior. We clarify here that the anticipatory locomotor adjustment is not only an adaptation of step width, but that speed and step length can also be modified \cite{ref65, ref68, ref66, ref67, ref61}.

In virtual reality experiments, participants are generally asked to wear a motion capture system and HMD and to perform locomotion sequences while avoiding the virtual content. Several experiments \cite{ref17, ref7} have focused on collision avoidance with objects (e.g., cylinders) instead of virtual characters. However, collision avoidance between participants and virtual characters has also been examined. By studying the human-virtual human avoidance behavior, Bailenson et al. \cite{ref12} shown that participants maintained a greater distance from virtual humans when walking toward them from their fronts compared with their backs. Comparisons of collision avoidance trajectories in real and virtual conditions in a collision avoidance task were also conducted by Olivier et al. \cite{ref69}. The experiment conducted by Bonsch et al. \cite{ref22} concluded that participants preferred collaborative collision avoidance (they expected the virtual character to step aside to get more space to pass but were willing to adapt their own walking paths) in small-scale virtual environments. Cinelli et al. \cite{ref68} studied the distance at which the participants start to deviate from their initial path. Sanz et al. \cite{ref5} investigated obstacle avoidance behavior during real walking in a large immersive projection set-up by analyzing the walking behavior of participants when avoiding real and virtual static obstacles. To generalize their study, they considered both anthropomorphic and inanimate objects. Their results showed that participants exhibit different locomotion behaviors in the presence of real and virtual obstacles and in the presence of anthropomorphic and inanimate objects. Finally, a study on collision avoidance using an HMD was conducted by Silva et al. \cite{ref70} to understand the way participants interact with virtual characters in immersive virtual environments.

\subsection{Gaze and Interactions}
\label{sec22}
The effects of gaze interaction between users and virtual characters has also been examined in the past. It has been found that during interactions with humans, the gaze \cite{ref71} and mutual eye contact \cite{ref72} can be interpreted as a core social interaction mechanism and main social interaction factor.

In addition, gaze interaction has been examined during walking tasks. The study conducted by Bailenson et al. \cite{ref74} indicated that more personal space was given to virtual characters by the users who engaged in mutual gaze. Narang et al. \cite{ref75} found that the gaze of a virtual character toward a walking user improved the sense of immersion. Nummenmaa et al. \cite{ref76} found that participants used their gaze as a cue to avoid collision by changing their path to the opposite side of the character's gaze. Finally, the virtual reality study conducted by Lynch et al. \cite{ref73} examined the effect of gaze interception during collision avoidance between two walkers. The authors concluded that the mutual gaze can be considered as a form of nonverbal communication between participants and virtual characters.

\subsection{Locomotion Analysis}
\label{sec23}
Analyzing human locomotion has been an extensive field of study not only for virtual reality, but also for kinesiology researchers, who have studied and proposed different ways of analyzing participants' locomotive behavior. In most cases, criteria related to task completion time, traveled distance, number of collisions, and path precision with respect to the ideal path have been used in a variety of studies \cite{ref59, ref60, ref58, ref57, ref34}. An alternative method includes the empirical observations of trajectory visualizations \cite{ref45}.

A number of studies have used distance metrics between trajectories \cite{ref56, ref54, ref55}. Fink et al. \cite{ref17} proposed a set of metrics, namely the mean radius of curvature along the full path, the maximum Euclidean distance from a straight line between the origin and the target, and the minimum Euclidean distance between the path and the obstacles of the virtual environment. Principal component analysis of a set of trajectories has also been used \cite{ref32}. The stride length, step width, variability in stride velocity, and variability in step width have also been used to evaluate and compare trajectories generated in virtual and real environments based on the gait cycle of walkers \cite{ref41, ref57}. Finally, Cirio et al. \cite{ref4} proposed nine metrics related to the shape, performance, and kinematic features that could be used to compare virtual and real trajectories. To evaluate the participants' trajectories in the current study, we adopted metrics proposed in Cirio et al. \cite{ref4}.

\section{Methodology and Implementation}
\label{sec3}
\subsection{Participants}
\label{sec31}
We conducted an a priori power analysis to determine the sample size, using G$^*$Power 3.1.9.2 software \cite{ref77}. The calculation was based on 95\% power, a medium-effect size of 0.25 \cite{ref78} with four repeated measures, a non-sphericity correction $\epsilon = 0.60$, and an $\alpha = 0.05$. The analysis resulted in a recommended sample size of 52 participants.

The participants were recruited in various ways: posters on campus, e-mails throughout the departments of the University, and in-class announcements. Participants provided informed consent in accordance with the Institutional Review Board of the University of the Aegean. No direct reward was given for participation, but three participants received a \euro{30} gift card after a lottery.

Initially, 55 students volunteered to participate in the experiment. Of them, two students did not follow the experimenter's guidelines properly. Another student identified himself as having chronic conditions that might affect his locomotion behavior. It is clarified that these students took part in our experiment; however, we did not use the collected data to analyze their collision avoidance behavior.

The final sample used in analyses consisted of 52 participants. Of these, 34 were male and 18 were female. Their ages varied from 19 to 33 ($M = 23.15$, $SD = 2.72$).

\subsection{Conditions of the Experiment}
\label{sec32}

\begin{figure}[htb]
\centering
\includegraphics[width=\columnwidth]{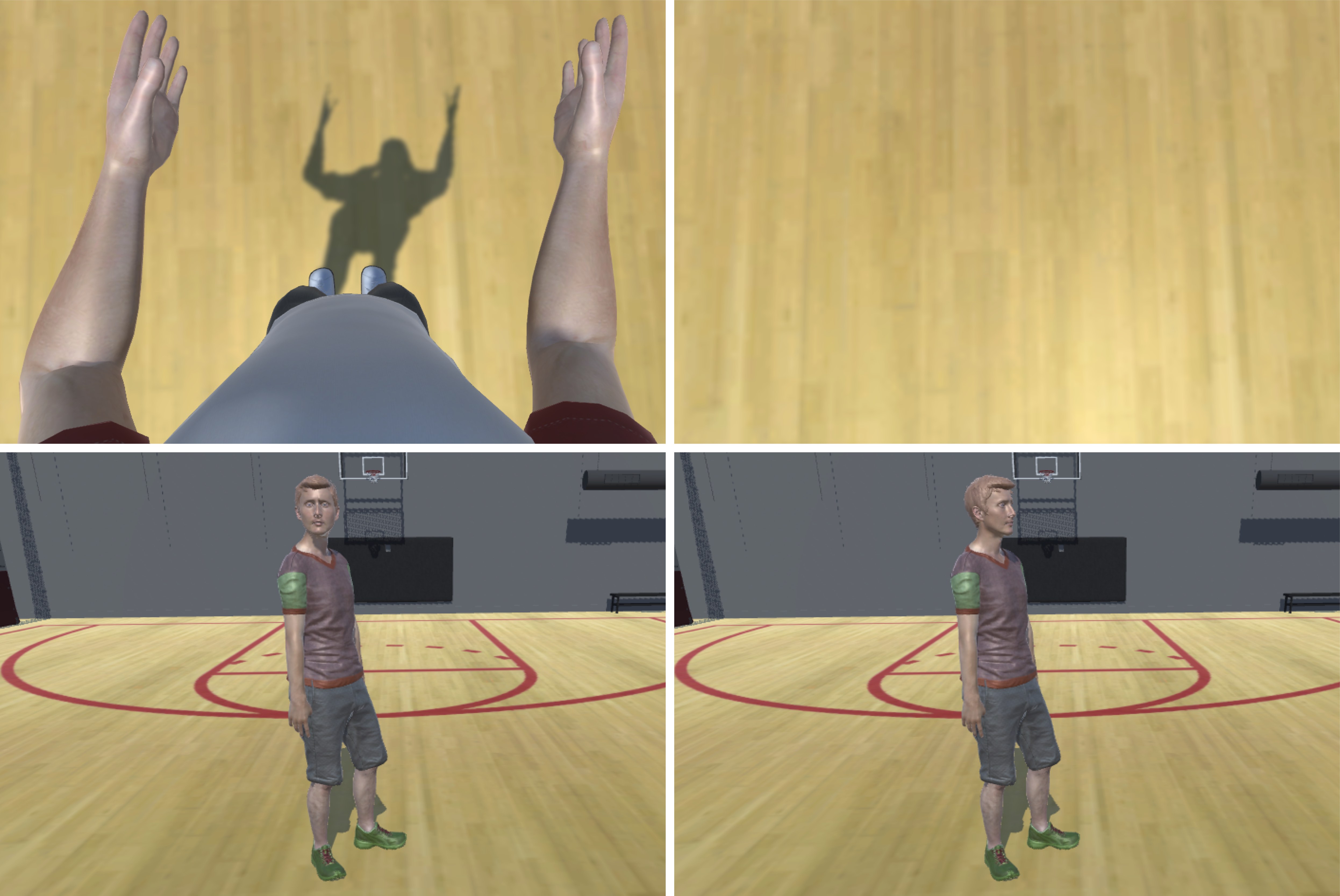}
\caption{The two dimensions of the experiment. Top: Self-Avatar versus No Self-Avatar. Bottom: LookAt versus No LookAt.}
\label{fig2}
\end{figure}

Within the four experimental conditions we can distinguish two different dimensions which are shown in Figure \ref{fig2}: Self-Avatar versus No Self-Avatar (conventionally Avatar dimension) and LookAt versus No LookAt (conventionally Gaze dimension). The four conditions were: (a) Self-Avatar LookAt, (b) No Self-Avatar LookAt, (c) Self-Avatar No LookAt, and (d) No Self-Avatar No LookAt. Its condition combine two situations. That is, during the Self-Avatar situation a virtual avatar is used to represent participants within the virtual world. During the No Self-Avatar situation, no virtual avatar is used, therefore the participant's body is invisible. During the LookAt situation, the character's gaze follows the participant. Finally, during the No LookAt situation, the idle motion of the character is used to animate the head of the virtual character. Based on the answers of the participants, we were able to configure the environment to assign the appropriate gender to the self-avatar so that it could represent the participant and avoidance character.

\subsection{Experiment Procedure}
\label{sec33}
The participants came to the location at which the experiment was conducted. The experimenter informed them about the project, and participants were briefly introduced to the motion capture system and the virtual reality headset. Participants were then asked to wear the motion capture system and the HMD, and to walk in the virtual environment to ensure they felt comfortable with these devices.

The virtual environment in which the participants were asked to walk was the same one used in our experiment; however, no content (virtual character or marks on the floor) related to the experiment were present during the practice walks. After becoming acquainted with the virtual reality equipment, the participants were asked whether they felt comfortable and were ready to participate in the experiment. The structure of the experiment was explained to them, but the specific virtual reality conditions that the participant would face were not mentioned. The participants were informed that they were allowed to have short breaks between the conditions and that they were allowed to quit the experiment at any point without any repercussions. The total duration of the procedure lasted on average 30 minutes.

During the experiment, the participants were asked to walk from the starting position to a target position that was visible in the virtual environment for all four experimental conditions (Self-Avatar LookAt, No Self-Avatar LookAt, Self-Avatar No LookAt, No Self-Avatar No LookAt). Each participant performed each experimental condition once. The sequence of the experiment was randomized, ensuring counterbalancing between conditions. Finally, an in-app countdown signal was used to inform participants when they should start walking.

\subsection{Equipment, Application, and Implementation}
\label{sec34}
For this study, the devices used were the Oculus Rift HMD with the TPCast wireless adapter for projecting the virtual reality content and the Perception Neuron for capturing the motions of the participants. The motion capture system transmitted the captured data wirelessly. This ensured that the participants were able to walk freely, since no wires were used. A participant equipped with the motion capture system and the HMD is shown in Figure \ref{fig1}.

\begin{figure*}[htb]
\centering
\includegraphics[width=\textwidth]{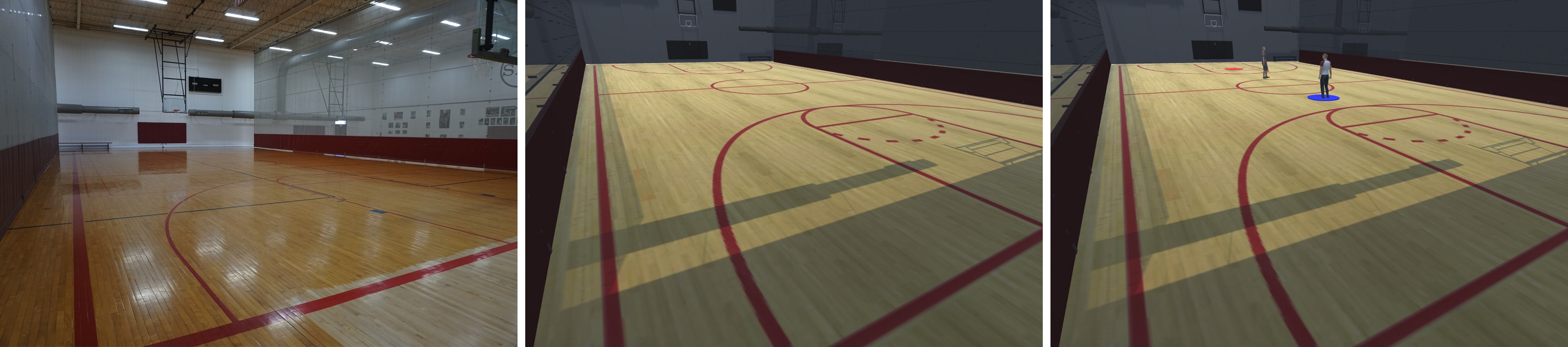}
\caption{The basketball court reserved for our study (left), the empty virtual environment to which the participants were first introduced (center), and the virtual environment (application scene) in which the experiment took place (right).}
\label{fig4}
\end{figure*}

The application used in the experiment was developed in Unity3D. The applications consisted of a single scene, as shown in Figure \ref{fig4}, which contained the main virtual reality application that we developed. A blue and red indicator (cycle) informs the participants about the starting (blue indicator) and target (red indicator) positions. At the midway point, a virtual character is placed. The gender of the virtual character is configured based on the gender of the participant. This is the character that the participant should avoid. This character was assigned an idle motion. For the LookAt dimension, the Unity3D built-in function of LookAt (\texttt{transform.LookAt[target]}) was used to rotate the head of the virtual character in a way that followed the global position of the user. An invisible collider was placed on the boundaries of the starting position indicator to detect whether the participant exited the starting area. Similarly, an invisible collider was placed on the target indicator to detect when the participant entered the target position. Collision detection was used to start the motion capture process when the participant exited the collider at the starting position, to stop the recording when the participant entered the collider at the target position, and to save the recording when the participant entered the starting position (going back to continue with the experiment). The experimenter was able to inspect and control this process by using Unity3D inspector.

During intervals between the conditions of the experiment, a black screen was shown to participants. We decided to add the black screen in between the conditions to avoid the virtual character appearing in a non-natural way (appearing out of thin air, violating laws of physics). This ensured that the manner in which the characters appeared did not affect participants' reactions. Finally, to ensure that all captured motion sequences were spatially aligned, a simple method was developed that placed the target position at the exact distance from the participant's position and direction (we used the forward vector of the participant to find the exact forward position at which the target indicator should be placed) and that placed the avoidance character in the appropriate position. This process was controlled by the experimenter using a button in Unity3D inspector.

The virtual reality experiment was conducted at the University of the Aegean recreation center on a basketball court that was reserved for 3 weeks. The dimensions of the court are 28 meters long and 15 meters wide. These measures were used to approximate the dimension of the virtual space. The court was free of objects and obstacles, making it ideal for conducting this type of experiment. A simple virtual environment (see Figure \ref{fig4}) was designed in 3ds Max and imported to a Unity3D game engine to approximate the physical space of the room. 

\begin{figure}[htb]
\centering
\includegraphics[width=0.495\columnwidth]{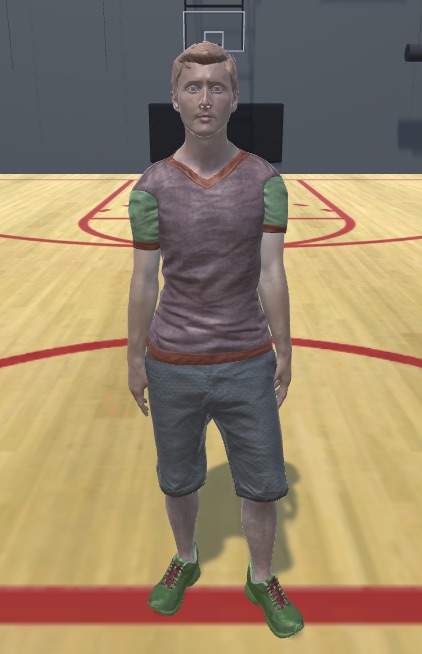}
\includegraphics[width=0.495\columnwidth]{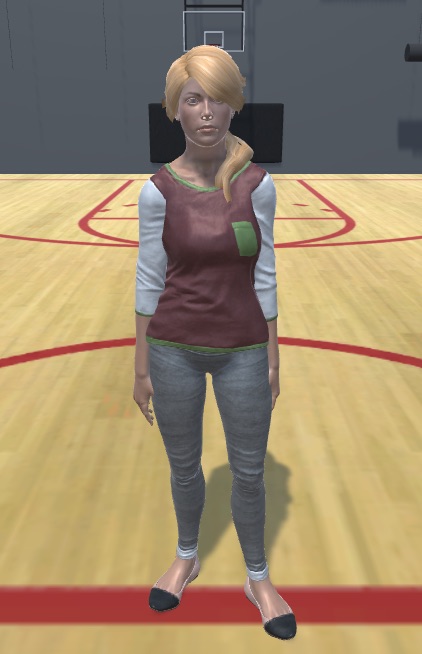}
\caption{The characters used in our experiment that participants were asked to avoid. The assigned gender of the avoidance character matched the gender of the participant.}
\label{fig5}
\end{figure}

The virtual characters used in the study were designed using Adobe's Fuse software. Specifically, we designed two male and two female characters that were used as the self-avatars to represent the participants in the virtual environment, and the other two were used to represent the characters that the participants were asked to avoid. The characters the participants were asked to avoid are seen in Figure \ref{fig5}. It is noted that the avoidance character had the same gender as the self-avatar (participant-controlled character) because we wanted to avoid potential dissonance between different genders by capturing the avoidance behavior of participants when avoiding a virtual character of the same gender. If we had an additional character of the opposite gender, this would significantly increase the total duration of the experiment, and it might become boring for the participants to conduct the experiment, causing a loss of motivation. Thus, we decided to examine the influence of the virtual character's gender on avoidance behavior in future research.

To provide the participants with enough space and time to avoid the virtual character located midway, a total distance of 8 meters between the starting and target positions was chosen. In reaching our decision, we took into consideration (a) the previous study of Olivier et al. \cite{ref61} showing that avoidance maneuvers can start when the distance between two people is greater than 2 meters, (b) the work of Ducourant et al. \cite{ref62} on constant distance regulations, and (c) the Hicheur et al. \cite{ref63} study on interactions related to turns. Based on this literature, to capture smooth motions without sudden changes in the participants' trajectories, we decided that the walking motion of the participant should begin and finish 4 meters from the virtual character. In this way, no sudden changes and turns would be performed by the participants.

\subsection{Kinematic Measurements}
\label{sec35}
To understand the way in which the participants' trajectories changed when they avoided a character in the four conditions of the experiment, a set of quantitative measurements were adopted using those in Cirio et al.'s \cite{ref4} study. We also followed the methodology of Simeone et al. \cite{ref42}, and we filtered the captured trajectories in 100 equidistant points. The measurements that are listed below were calculated on the extracted points. Here, we isolated the trajectories of the participant route from the full-body motion that was captured. The kinematic measurements provided information about the motion and shape of the trajectories performed by the participants, which were used to represent the captured motion. The kinematic measurements were computed for each separate captured trajectory as follows:
\begin{enumerate}
\item \textbf{Length:} The total distance (length of the captured trajectory in centimeters) covered by the participants.
\item \textbf{Duration:} The total time (in seconds) that the participants needed to walk from the start to the target position.
\item \textbf{Speed:} The average speed (in centimeters/second) that the participants used to walk from the starting position to the target position.
\item \textbf{Deviation:} For each of the filtered points, the absolute value of the perpendicular distance to the closest segment was taken (in centimeters). Then, the average deviation for all 100 points is computed. This shows the deviation of the captured trajectory from the straight line. 
\end{enumerate}

\section{Results}
\label{sec4}
\subsection{Data Analysis}
\label{sec41}

\begin{table*}[htb]
\caption{Descriptive statistics (Mean [$M$], Standard Deviation [$SD$], Minimum [$Min$] and Maximum [$Max$] value) for each kinematic measure across experimental conditions ($N = 52$), and patterns of differences.}
\label{tab1}
\scriptsize%
\centering%
\begin{tabular}{p{4cm}>{\raggedleft\arraybackslash}p{1.7cm}>{\raggedleft\arraybackslash}p{1.7cm}>{\raggedleft\arraybackslash}p{1.7cm}>{\raggedleft\arraybackslash}p{1.7cm}l}
\toprule
\textbf{Experimental Condition} & \it{\textbf{M}} & \it{\textbf{SD}} & \it{\textbf{Min}} & \it{\textbf{Max}} & \textbf{Pattern of Differences} \\
\midrule
& \multicolumn{4}{c}{\textbf{Length (in cm)}}		&\\
\midrule
a. Self-Avatar LookAt		&973.83	&19.83	&923.84	&1016.62	&\multirow{4}{*}{}a $>$ c $>$ b $=$ d\\
b. No Self-Avatar LookAt		&914.81	&38.52	&838.10	&989.60	&\\
c. Self-Avatar No LookAt		&929.99	&23.74	&863.02	&972.64	&\\
d. No Self-Avatar No LookAt	&901.11	&22.92	&856.17	&943.45	&\\
\midrule
& \multicolumn{4}{c}{\textbf{Duration (in sec)}}		&\\
\midrule
a. Self-Avatar LookAt		&14.00	&.80		&11.58	&15.95	&\multirow{4}{*}{}a $>$ c $>$ d\\
b. No Self-Avatar LookAt		&13.38	&1.00	&12.15	&15.64	&b $=$ c\\
c. Self-Avatar No LookAt		&13.57	&0.99	&11.86	&15.97	&b $=$ d\\
d. No Self-Avatar No LookAt	&12.93	&1.09	&10.22	&15.91	&\\
\midrule
& \multicolumn{4}{c}{\textbf{Speed (in cm/sec)}}	&\\
\midrule
a. Self-Avatar LookAt		&69.76	&3.68	&61.17	&81.33	&\multirow{4}{*}{}a $=$ b $=$ c $=$ d\\
b. No Self-Avatar LookAt		&68.54	&5.08	&58.70	&80.08	&\\
c. Self-Avatar No LookAt		&68.83	&4.02	&58.31	&77.11	&\\
d. No Self-Avatar No LookAt	&69.76	&4.49	&59.26	&81.26	&\\
\midrule
& \multicolumn{4}{c}{\textbf{Deviation (in cm)}} &\\
\midrule
a. Self-Avatar LookAt		&54.82	&12.78	&32.69	&78.16	&\multirow{4}{*}{}a $>$ c $>$ b $=$ d\\
b. No Self-Avatar LookAt		&34.78	&9.15	&16.49	&47.78	&\\
c. Self-Avatar No LookAt		&41.78	&8.85	&25.45	&56.47	&\\
d. No Self-Avatar No LookAt	&33.62	&10.19	&17.99	&48.49	&\\
\bottomrule
\end{tabular}
\end{table*}

Our intent was to explore whether the kinematic data and trajectories differed in each kinematic measure across the four experimental conditions (Self-Avatar LookAt, No Self-Avatar LookAt, Self-Avatar No LookAt, No Self-Avatar No LookAt; see Research Questions 1-4). It should be noted that the length, duration, speed, and deviation represent different but related kinematic mesures, and are also expressed in different units of measurement (centimeters, seconds, and centimeters/second) that result in very different variances. Therefore, we conducted four separate one-way repeated-measures analysis of variance (ANOVA) corresponded to each of the four experimental conditions with four dependent variables corresponded to each kinematic measure (length, duration, speed, and deviation).

We screened variables for univariate and multivariate normality, linearity among variables, outliers, equality of error variances for each of the dependent variables, and sphericity. All these assumptions were met \cite{ref79}. For the statistical analyses, the IBM SPSS Statistics software v. 23.0 was used \cite{ref1}.

\begin{figure*}[htb]
\centering
\includegraphics[width=0.4975\textwidth]{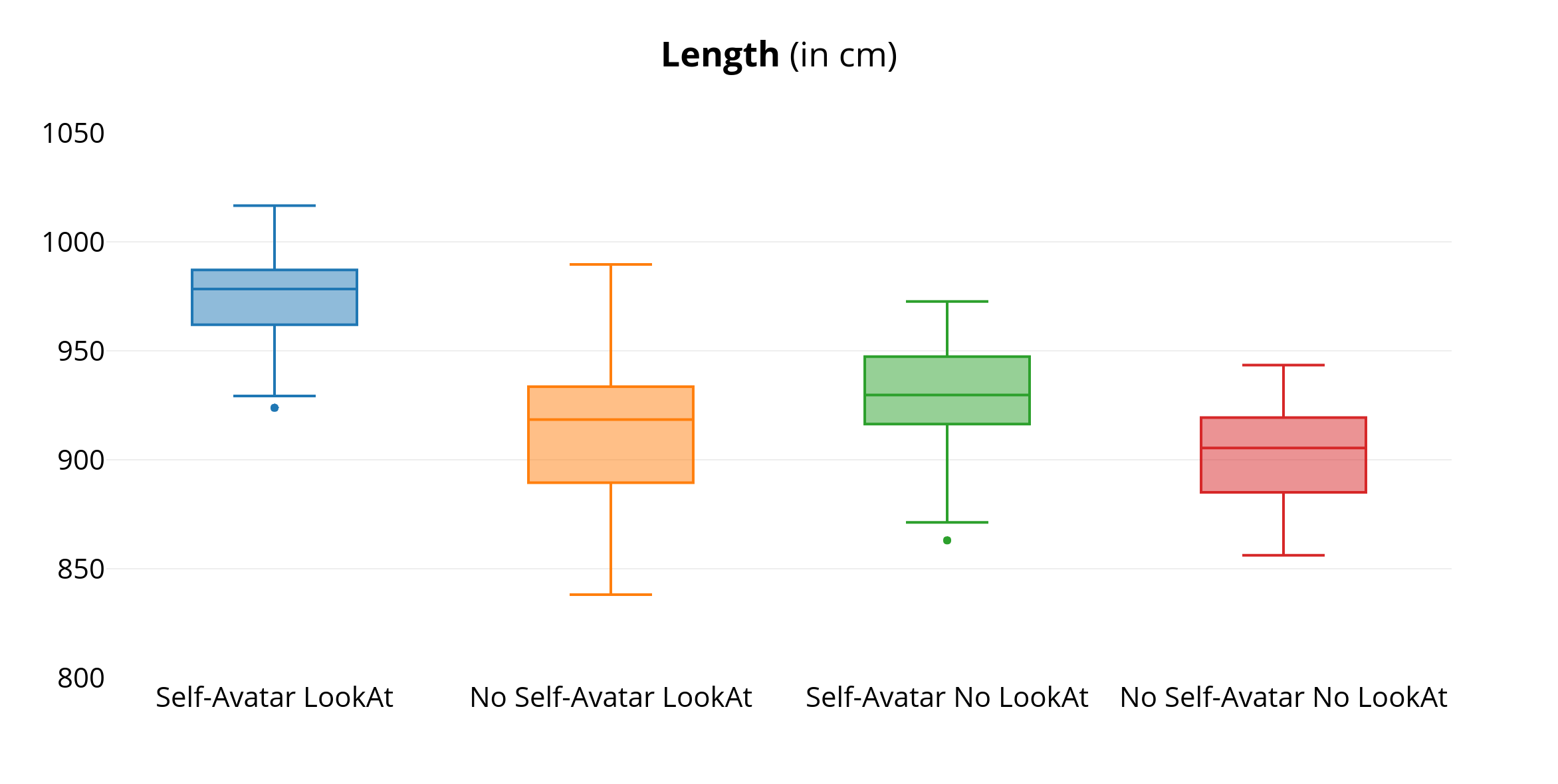}
\includegraphics[width=0.4975\textwidth]{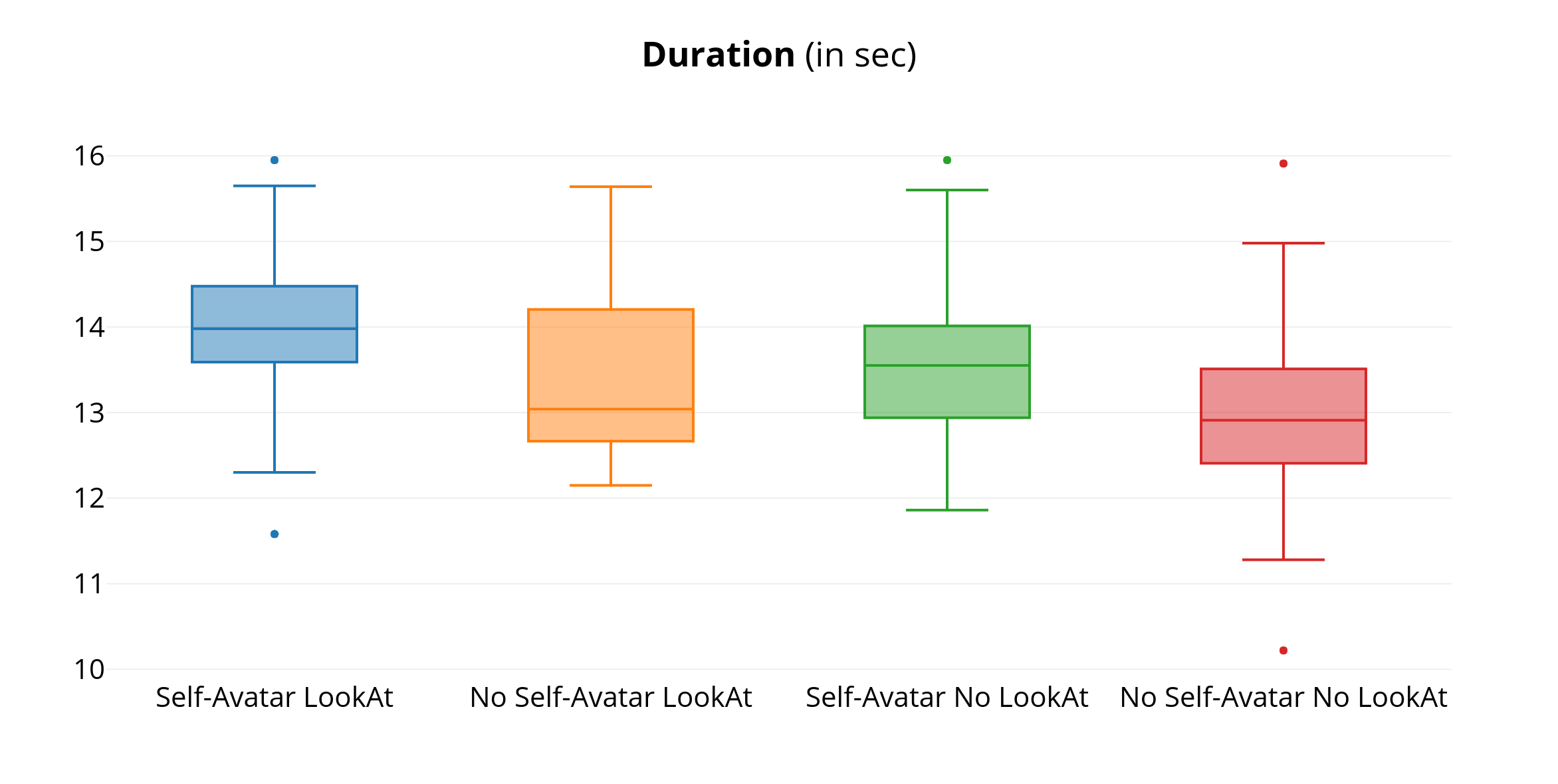}
\includegraphics[width=0.4975\textwidth]{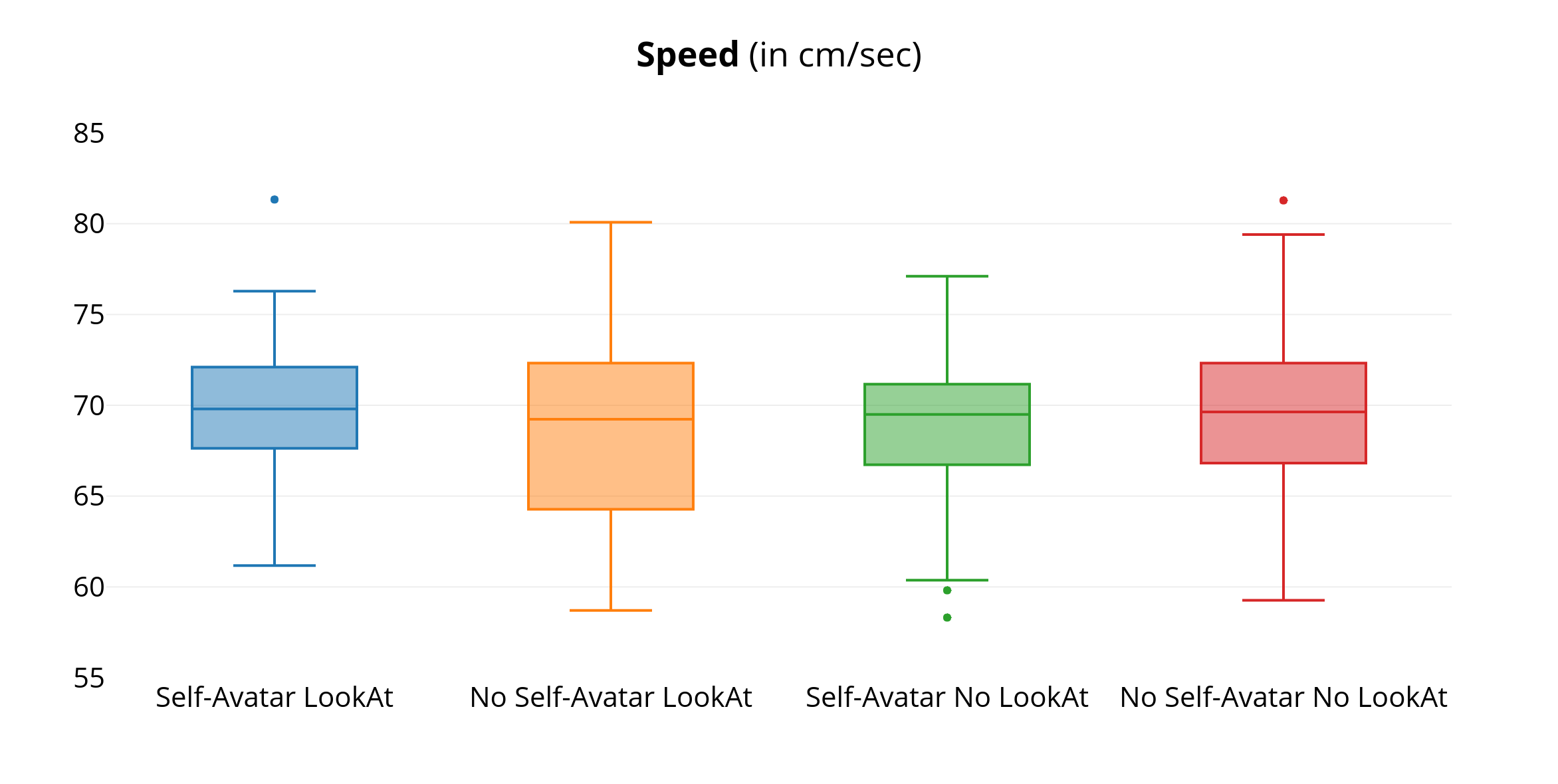}
\includegraphics[width=0.4975\textwidth]{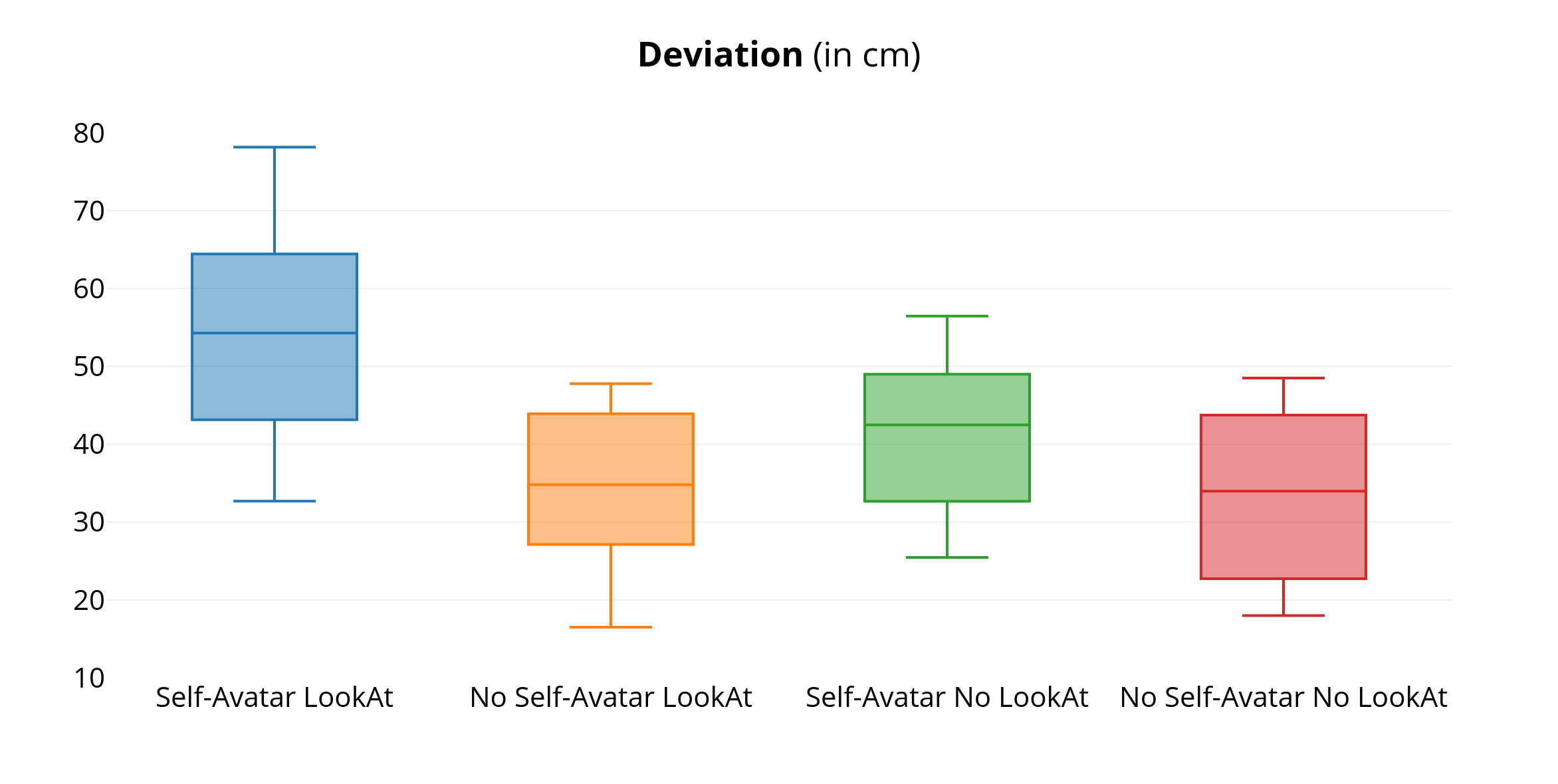}
\caption{Kinematic results for each kinematic variable. Boxes enclose the middle 50\% of the data. The median is denoted by a thick horizontal line. See Table \ref{tab1} for means and standard deviations.}
\label{fig6}
\end{figure*}

Table \ref{tab1} shows the descriptive statistics (means, standard deviations, minimum and maximum values) for each kinematic measure across the four experimental conditions. In addition, Figure \ref{fig6} indicates boxplots of kinematic measures in four panels for each kinematic measure (length, duration, speed, and deviation).

\subsection{Kinematic Analysis}
\label{sec42}
\begin{figure*}[tb]
\centering
\includegraphics[width=\textwidth]{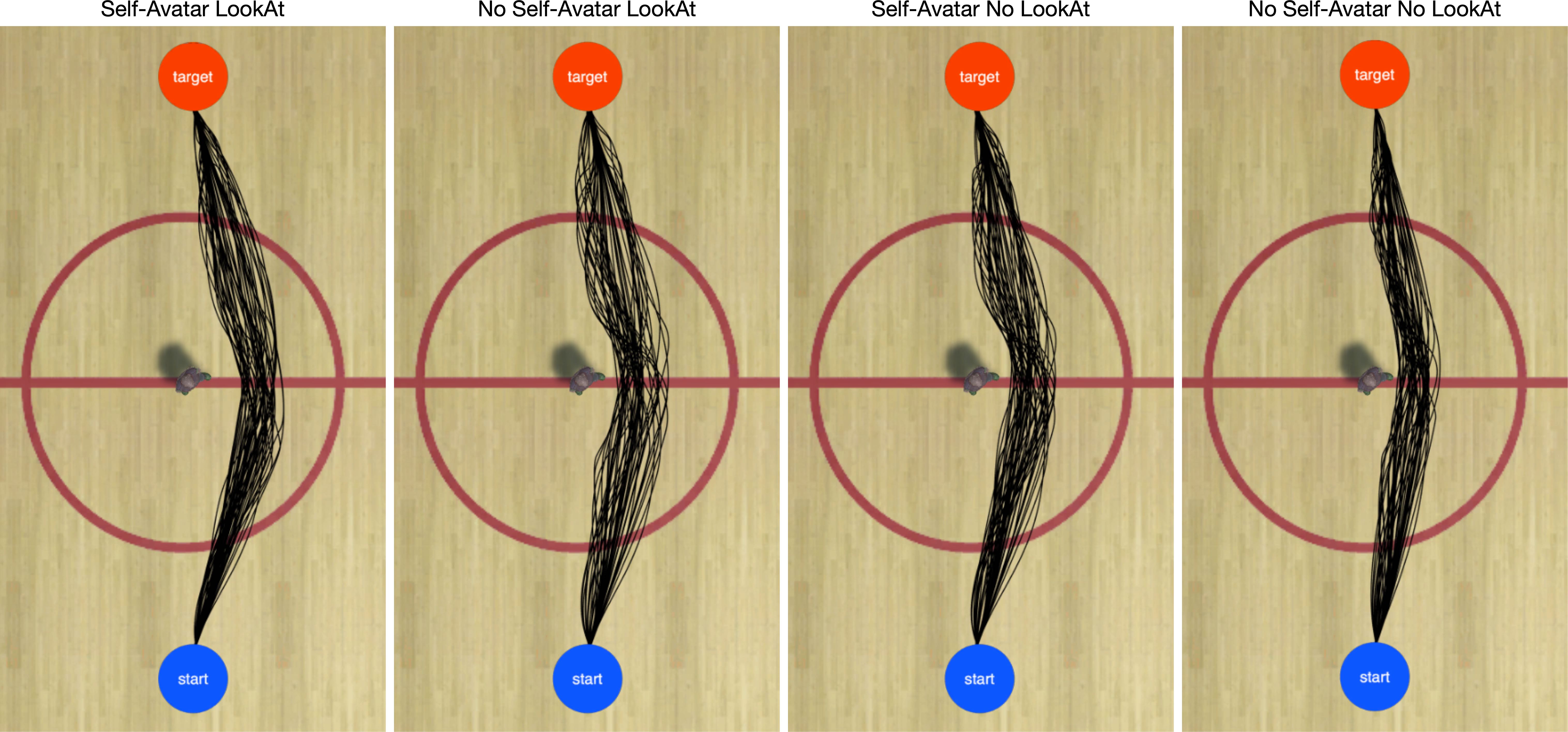}
\caption{The trajectories the participants followed for each condition of the experiment.}
\label{fig7}
\end{figure*}

Figure \ref{fig7} illustrates the path the participants followed for all four conditions. Note that all left-side avoidance trajectories mirrored to face the right side of the avoidance character for visualization purposes. For the \textbf{length} variable, the one-way repeated measures ANOVA with Greenhouse-Geisser-corrected estimates of sphericity ($\epsilon = .66$) revealed a significant difference: $F (1.98, 101.44) = 119.05$, $p<0.001$, $\eta_p^2=0.70$. Pairwise comparisons, using the Holm's sequential Bonferroni correction for controlling for Type I errors \cite{ref2}, indicated that participants during the Self-Avatar LookAt condition followed significantly longer paths than the No Self-Avatar LookAt condition ($p<0.001$), Self-Avatar No LookAt condition ($p<0.001$), and No Self-Avatar No LookAt condition ($p<0.001$). The path was significantly shorter in the No Self-Avatar LookAt condition compared with that in the Self-Avatar No LookAt condition ($p<0.05$), but not significantly different from the path in the No Self-Avatar No LookAt condition ($p=0.05$). Finally, the path in the Self-Avatar No LookAt condition was significantly longer compared with that in the No Self-Avatar No LookAt condition ($p<0.001$). In brief, the length pattern across the four conditions is as follows: 
\begin{itemize}
\item Self-Avatar LookAt $>$ Self-Avatar No LookAt $>$ No Self-Avatar LookAt $=$ No Self-Avatar No LookAt.
\end{itemize}

For the \textbf{duration} variable, the one-way repeated measures ANOVA with Greenhouse-Geisser-corrected estimates of sphericity ($\epsilon = .84$) revealed a significant difference: $F (2.53, 129.07) = 16.62$, $p<0.001$, $\eta_p^2=0.25$. Pairwise comparisons, using the Holm's sequential Bonferroni correction, indicated that the duration of the collision avoidance task during the Self-Avatar LookAt condition was longer than the task duration in the No Self-Avatar LookAt condition ($p<0.01$), Self-Avatar No LookAt condition ($p<0.05$), and No Self-Avatar No LookAt condition ($p<0.001$). No significant differences were found between the No Self-Avatar LookAt and Self-Avatar No LookAt conditions ($p>0.05$), as well as between the No Self-Avatar LookAt and the No Self-Avatar No LookAt condition ($p=0.05$). Finally, the task duration in the Self-Avatar No LookAt condition was significantly longer than that in the No Self-Avatar No LookAt condition ($p<0.001$). In brief, the duration pattern across the four conditions is as follows:
\begin{itemize}
\item Self-Avatar LookAt $>$ Self-Avatar No LookAt $>$ No Self-Avatar No LookAt, 
\item No Self-Avatar LookAt $=$ Self-Avatar No LookAt, and
\item No Self-Avatar LookAt $=$ No Self-Avatar No LookAt.
\end{itemize}

As for the variable of participants' average \textbf{speed}, the one-way repeated measures ANOVA with Greenhouse-Geisser-corrected estimates of sphericity ($\epsilon = .96$) revealed no significant differences: $F (2.26, 115.27) = 1.38$, $p>0.05$, $\eta_p^2=0.03$. In brief, the speed pattern across the four conditions is as follows: 
\begin{itemize}
\item Self-Avatar LookAt = No Self-Avatar LookAt = Self-Avatar No LookAt = No Self-Avatar No LookAt.
\end{itemize}

For the \textbf{deviation} variable, the one-way repeated measures ANOVA with Greenhouse-Geisser-corrected estimates of sphericity ($\epsilon = .91$) revealed a significant difference: $F (2.72, 138.84) = 45.81$, $p<0.001$, $\eta_p^2=0.47$. Pairwise comparisons, using the Holm's Ssquential Bonferroni correction, indicated that the deviation of the trajectory from the straight line during the Self-Avatar LookAt condition was larger than that in the No Self-Avatar LookAt condition ($p<0.001$), Self-Avatar No LookAt condition ($p<0.001$), and No Self-Avatar No LookAt condition ($p<0.001$). The deviation was significantly smaller in the No Self-Avatar LookAt condition compared with that in the Self-Avatar No LookAt condition ($p<0.001$). No significant differences were found between the No Self-Avatar LookAt condition and the No Self-Avatar No LookAt condition ($p>0.05$). Finally, the deviation of the trajectory from the straight line in the Self-Avatar No LookAt condition was significantly larger than that in the No Self-Avatar No LookAt condition ($p<0.01$). In brief, the deviation pattern across the four conditions can be represented as follows: 
\begin{itemize}
\item Self-Avatar LookAt $>$ Self-Avatar No LookAt $>$ No Self-Avatar LookAt $=$ No Self-Avatar No LookAt.
\end{itemize}

\section{Discussion }
\label{sec5}
The present study was based on four experimental conditions (Self-Avatar LookAt, No Self-Avatar LookAt, Self-Avatar No LookAt, No Self-Avatar No LookAt), using kinematic data, to investigate the effects of self-avatar and gaze of virtual character on users' movement behavior in a collision avoidance task. In our experiment, participants were embodied using a self-avatar.

Remarkably, we found that the length and duration of the participants' trajectory, as well as the deviation of the trajectory from the straight line, were greater in the Self-Avatar LookAt condition. Regarding the variable of average speed, no significant differences were found among participants across the four experimental conditions. Based on the kinematic measurements, we found that there are distinct differences between the Self-Avatar and the No Self-Avatar conditions. 

Overall, the patterns of statistically significant differences show (see Table \ref{tab1}) that the length and duration of the task, as well as the deviation of the captured trajectories from the straight line, were greater in the Self-Avatar represented conditions. Moreover, in the LookAt conditions, the overall pattern shows an effect of gaze on length, duration, and deviation. Specifically, the gaze seems to have an additional effect on the spatial measures (length and deviation), as the above combination of the patterns shows.

Although we found statistically significant differences in the duration measure across the conditions, no such differences were found in the average speed (in centimeters/second) of the participants. It is very interesting that an inspection of the relationship between duration and speed in each condition revealed high bivariate correlations (Pearson $rs$ were $-.94$, $-.83$, $-.91$, $-.89$ across the four conditions). In addition, the bivariate correlations between duration and length were low to moderate ($rs$ were $.40$, $.27$, $.48$, $.37$) and negligible to low between duration and deviation ($rs$ were $.07$, $-.05$, $-.25$, $.13$). The bivariate correlations between speed and length were negligible to low ($rs$ were $-.08$, $.30$, $-.13$, $-.12$), similarly low to negligible between speed and deviation ($rs$ were $-.07$, $.02$, $.25$, $-.15$). Finally, the bivariate correlations between length and deviation were negligible ($rs$ were $-.00$, $-.04$, $-.06$, $-.07$). Taking the results of the repeated ANOVAs and the aforementioned patterns of correlations together, we can infer that the length of the trajectories was the main factor that critically influenced the duration of the task, while the speed of participants tended to be steady. The participants seemed to regulate their collision avoidance behavior mainly in terms of length rather than deviation, whereas the average speed remained the same across the four conditions when facing a rather normal virtual character in this experiment (not an attractive or repellant character). In general, the four conditions of the experiment had greater differentiated impact on length of the trajectory and duration of the task.

Furthermore, the findings of the Self-Avatar situations indicate that the self-avatar assigned to participants had an impact on their collision avoidance behavior. It seems that our participants changed their paths and followed longer routes to ensure the avoidance of collisions with the virtual character. Thus, when participants are represented by a self-avatar, they might become aware of this avatar and the potential collisions that might happen, since they have volume/mass that the participants can observe in themselves when viewing the avatar. In other words, the Self-Avatar dimension may enhance the participants' sense of embodiment since participants have a body and are not invisible.

Another interpretation is that during the No Self-Avatar conditions, the participants felt less present and more invisible. Thus, the inability to observe a virtual body might also affect the way that the participants decided to avoid the virtual character. It is a common trope in a number of science fiction movies for invisible personas to be able to pass through objects, walls, and human bodies, and this prior knowledge might also have affected the path the participants chose to follow because they might have felt more comfortable passing closer to or through the virtual character. It should be noted that none of the participants in our experiment passed through the virtual character, although a number of them narrowly avoided the character. In any case, the aforementioned interpretations indicate that the use of a self-avatar to represent the participants in the virtual environment had an impact on the path they followed when avoiding the virtual character, making them more aware of the environment and the possible collisions that might happen.

For the LookAt (Gaze) situations, the path the participants chose had a significant impact, especially on length and duration, compared with the No LookAt situations, especially when combined with Self-Avatar situations. Observed by the virtual character continuously, the participants could sense that they were in a virtual environment. Our findings show that the presence of the participants in the LookAt conditions is greatly enhanced when their virtual body is observed by a virtual character. Conversely, the absence of a self-avatar made the LookAt situation have a small or negligible impact on kinematic measures.

A few issues raised after the end of this experiment need to be considered in future studies. We provide here some concerns followed by suggestions for improvement. The first concern is related to the adjustments made to the self-avatar to match the participant. Specifically, we did not resize the self-avatar to match a participant's height and weight. The participant's height was retargeted to match the height of the virtual character. The weight factors were not considered at all. We assume that the lack of customization of a character to match a participant's weight may have negatively affected our experiment. Another customization that was not employed was the skin color of the character. We chose to use a Caucasian color for the self-avatar due to the geographic location where the experiment was conducted. However, depending on the geographic location and the participant pool, self-avatars of other skin colors might be required, since the use of a self-avatar with a single light skin color might affect participants' presence and embodiment.

\section{Conclusions and Future Research}
\label{sec6}
Based on the analysis of the kinematic measures, we found that there were changes in the avoidance behavior of the participants. There were distinct differences in the paths that the participants choose to follow associated with presence and embodiment. Specifically, the results show that the Self-Avatar situations had the greatest impact on length, duration and deviation, followed by the LookAt situations. This means that when a self-avatar is assigned to a participant, the participant becomes more aware of the environment and potential collisions with the virtual character. A greater differentiated impact of the four conditions of the experiment was found for the length and duration of the task, whereas the average speed of participants tended to be steady.

Future research can investigate the ways that participants avoid virtual characters with different features or genders and based on different participants' ages. In addition, avoidance behavior of participants could be studied when they are instructed to avoid virtual characters with variations in their appearance and motion similarly to Mousas et al. \cite{ref28}. Because the current approach only examined static (only idle motion was applied) virtual characters, it is vital to further examine the interactions between locomotive virtual characters and understand the way that the participants interact with groups of virtual characters and virtual crowds.

The current study is a step toward understanding the presence and embodiment of participants during their avoidance behavior when interacting with virtual populations that have a variety of bodily and facial features. Future research can validate whether kinematic measures significantly correlate with presence and embodiment factors typically evaluated from self-reported data (questionnaires). Finally, we assume that a number of interesting insights about the participants' movements can be obtained when performing biomechanical analysis of the captured full-body motion of participants.

\bibliographystyle{abbrv-doi}

\bibliography{template}
\end{document}